\documentclass[final,5p,twocolumn,times]{elsarticle}

\usepackage{graphicx}
\usepackage{dcolumn}
\usepackage{bm}

\newcommand {\be}{\begin{equation}}
\newcommand {\ee}{\end{equation}}
\newcommand {\bea}{\begin{eqnarray}}

\newcommand {\eea}{\end{eqnarray}}

\title{Resonant formation of $\Lambda(1405)$ by stopped-$K^-$ absorption in deuteron
}

 \author[rvt,irn]{Jafar Esmaili\corref{cor1}
}
\ead{jesmaili@riken.jp}
\author[rvt,focal]{Yoshinori Akaishi
}
\ead{akaishi@post.kek.jp}
\author[rvt,els]{Toshimitsu Yamazaki
}
\ead{yamazaki@nucl.phys.s.u-tokyo.ac.jp}
\cortext[cor1]{Corresponding author; jesmaili@riken.jp}
 \address[rvt]{RIKEN, Nishina Center, Wako, Saitama 351-0198, Japan}
\address[irn]{Department of  Physics, Isfahan University of Technology, Isfahan 84156-83111, Iran}
\address[focal]{College of Science and Technology, Nihon University, Funabashi, Chiba 274-8501, Japan}
\address[els]{Department of Physics, University of Tokyo, Bunkyo-ku, Tokyo 113-0033, Japan}

\begin{document}

\begin{abstract}
   To solve the current debate on the position of the quasi-bound $K^-p$ state, namely, ``$\Lambda(1405)$ or $\Lambda^*(1420)$", we propose to measure the $T_{21} = T_{\Sigma\pi \leftarrow \bar{K}N}$ $\Sigma\pi$ invariant-mass spectrum in stopped-$K^-$ absorption in deuteron, since the spectrum, reflecting the soft and hard deuteron momentum distribution, is expected to have a narrow quasi-free component with an upper edge of $M = 1430$ MeV/$c^2$, followed by a significant ``high-momentum" tail toward the lower mass region, where a resonant formation of $\Lambda(1405)$ of any mass and width in a wide range will be clearly revealed. We introduce a ``deviation" spectrum as defined by $DEV$ = $OBS$ (observed or calculated) / $QF$ (non-resonant quasi-free), in which the resonant component can be seen as an isolated peak free from the $QF$ shape.

\end{abstract}

\maketitle

\section{Introduction}

Where is the position of the $I=0$ $L=0$ $K^-p$ quasi-bound state? This question is directly connected to the strength of the s-wave $I=0$ $\bar{K} N$ interaction. Traditionally, the $\Lambda(1405)$ resonance is identified to this state \cite{PDG,Dalitz-Tuan:60,Dalitz-Wong:67}, and a strongly attractive $\bar{K} N$ interaction is indicated, which is compatible with theoretical predictions based on meson-exchange \cite{Speth:90} and on chiral dynamics \cite{Waas:96}. The attractive $\bar{K}N$ interaction was evidenced by an anomalously large subthreshold production of $K^-$ in heavy-ion reactions \cite{GSI-K}. Starting from this $\Lambda(1405)$ ansatz, Akaishi {\it et al.}   constructed phenomenologically an energy-independent $\bar{K} N$ complex potential by a $\bar{K}N$-$\Sigma \pi$ coupled-channel procedure, and predicted deeply bound dense kaonic nuclear systems \cite{AY02PRC,AY02PLB,AY04PLBa,AY04PRC,AY04PLBb}. Recently, on the other hand, another theoretical framework of chiral dynamics including a double-pole hypothesis has been proposed, claiming that the $K^- p$ quasi-bound state is located around 1420 MeV or higher  \cite{Oller:01,ORB,JOORM,HNJH,BNW,BMN,HW}. We name such a hypothetical state ``$\Lambda^*(1420)$". The predicted regime leads to a much less-attractive $\bar{K} N$ interaction, and thus, only shallow bound states are expected. It is vitally important to distinguish between $\Lambda(1405)$ and $\Lambda^*(1420)$ experimentally, but there seem to be lots of confusing statements concerning the strategy as to how valid experimental evidence can be obtained.

To clarify the situation let us take a simple and sufficient coupled-channel regime with $\bar{K}N$ (=1) and $\Sigma\pi$ (=2)  channels, since these two play a dominant role in the coupled-channel dynamics, as shown by Hyodo and Weise \cite{HW}. The imaginary part of the transition matrix $T_{11} = T_{\bar{K}N \leftarrow \bar{K}N}$, which corresponds to the imaginary part of the forward scattering amplitude, has a peak with a mass $M$ (either 1405 or 1420 MeV/$c^2$), but, since it is located below the $\bar{K} + N$ binding threshold, it cannot be detected directly. On the other hand, the transition matrix $T_{21} = T_{\Sigma\pi \leftarrow \bar{K}N}$ is responsible for the observation of $\Sigma\pi$ pairs following $K^-$ capture (or equivalently, following the associated production together with $K^+$).

\begin{figure}[htb]
\begin{center}
\includegraphics[width=8cm]{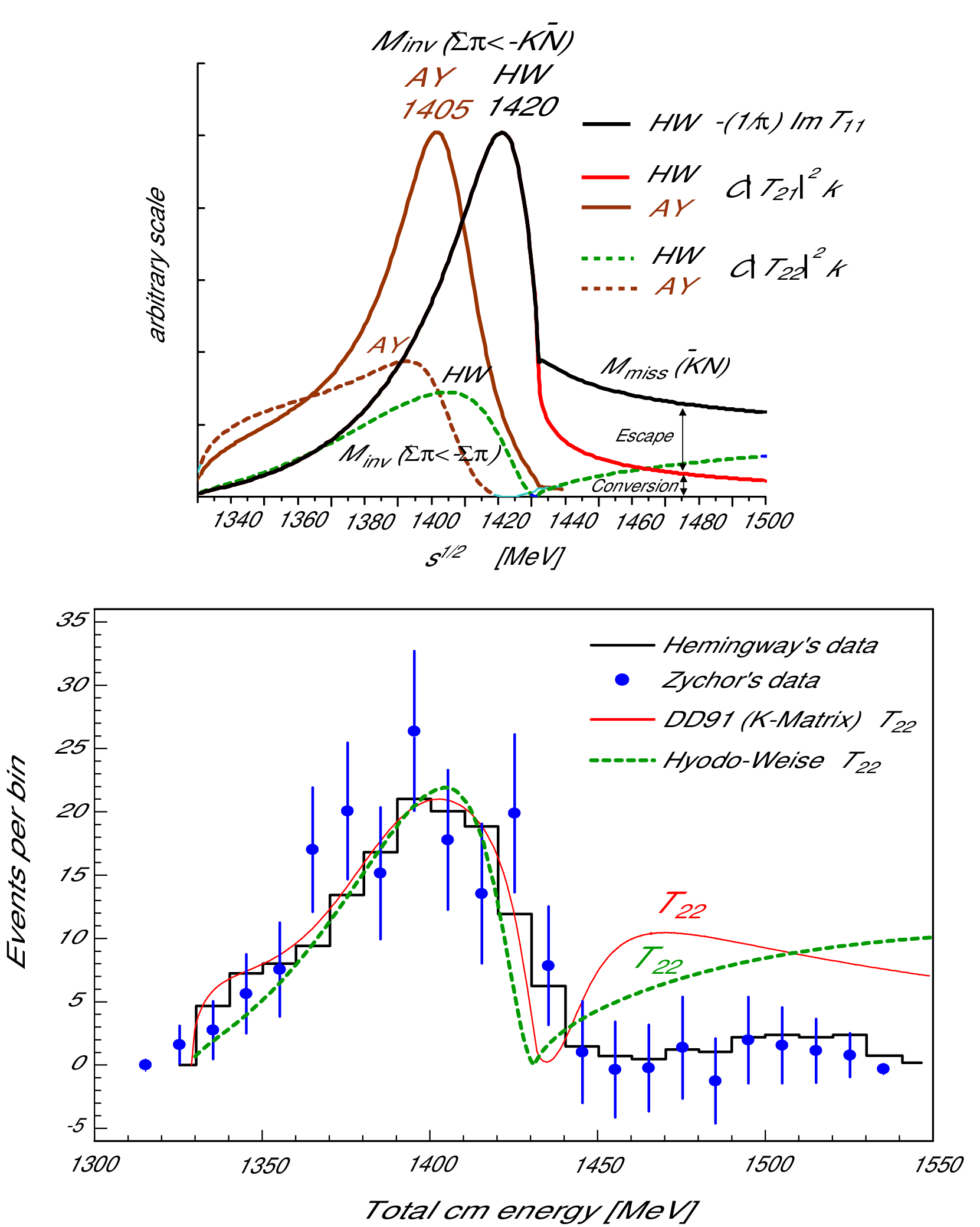}\\
  \caption{(Upper)  $(-1/\pi)\,{\rm Im} \,T_{11}$, $ | T_{21} |^2 k$ and $| T_{22} |^2 k$ curves in the chiral model of Hyodo-Weise \cite{HW}, in comparison with the $\Lambda(1405)$ ansatz of A-Y \cite{AY02PRC}. (Lower) The experimental $M_{\Sigma\pi}$ spectrum of Hemingway \cite{Hemingway85}, when fitted by $T_{22}$ of the
  phenomenological model of DD91 \cite{DD91} and the chiral model of HW. Zychor's data \cite{Zychor} are also shown.}\label{fig:MSigmaPi}
\end{center}
\end{figure}

Recent chiral theories predict two poles \cite{Oller:01,JOORM}. The 1st pole, mainly coupled to $\bar{K}N$, corresponds to a peak around 1420 MeV/$c^2$, as shown in Fig. 4 of \cite{HW}. (The 1st-pole position is even higher, around 1430 MeV, but its peak position is pushed down by the effect of the $\bar{K}N$ threshold). The 2nd pole, mainly coupled to $\Sigma\pi$, is broadly distributed with a non Breit-Wigner form in $T_{22} = T_{\Sigma \pi \leftarrow \Sigma \pi}$, and thus, {\it it cannot be detected directly} \cite{Akaishi:09}. It will be shown later that any $M_{\Sigma\pi}$ spectrum involving $T_{21}$ is dominated by the pole of $K^-p$, as Jido {\it et al.} \cite{JOORM} has already discussed; it is not much affected by the presence of a 2nd pole in the $\Sigma\pi$ channel. Thus, the issue of $\Lambda(1405)$ vs $\Lambda^*(1420)$ is a well-defined problem, and can be solved by experimental observables involving $T_{21}$.

In our recent work \cite{Esmaili} we have shown that the $\Sigma\pi$ invariant-mass spectra ($M_{\Sigma\pi}$) in stopped-$K^-$ absorption in $^4$He, $^3$He and $d$, which are governed by the spectator momentum distributions of $t$, $d$ and $n$, respectively, do reflect the resonant formation of a quasi-bound $K^-p$ state, contrary to the past interpretation in terms of the non-resonant direct-capture process. Thus, the issue of the location of the $K^-p$ quasi-bound state can be examined experimentally by a quantitative comparison of an observed $M_{\Sigma\pi}$ spectrum with predicted theoretical distributions including resonant formation. We made a $\chi^2$ analysis of  old bubble-chamber data of $M_{\Sigma\pi}$($K^- \, ^4$He) \cite{Riley} by varying the mass ($M$) and width ($\Gamma$) of an assumed resonance and found a significant minimum in the $M-\Gamma$ contour presentation of $\chi^2$ at
\be \label{eq:Riley-results}
M = 1405.5^{+1.4}_{-1.0}~{\rm MeV}/c^2~ {\rm and}~ \Gamma = 23.6 ^{+4}_{-3}~{\rm MeV}.
\ee
Thus, the $\Lambda^*(1420)$ ansatz is excluded by more than 99.9\% statistical confidence. However, since the $\Lambda(1405)$ signal does not appear as a separate peak, but as a small component involved in the steeply falling tail, we seek for a more convincing experimental method to isolate the resonance signal.

In the present paper, we point out that the use of a deuteron target in the reaction,
\begin{eqnarray}
{\rm stopped-}K^- + d \rightarrow &X&+ ~n,\\
&X&\rightarrow \Sigma + \pi,
\end{eqnarray}
can provide a more decisive information. Since the deuteron wavefunction is composed of low- and high-momentum components, the dominant ``quasi-free (QF)'' shape of $M_{\Sigma\pi}$ is narrow enough so that the resonant formation of even $\Lambda^*(1420)$ may be observable as a bump in the falling QF peak, whereas its tail, resulting from the high-momentum component of $d$, extends to the region where a resonant formation of $\Lambda(1405)$ can be revealed as a separate peak. We investigate this problem in detail.

\begin{figure}[htb]
\begin{center}
 \includegraphics[width=7.5cm]{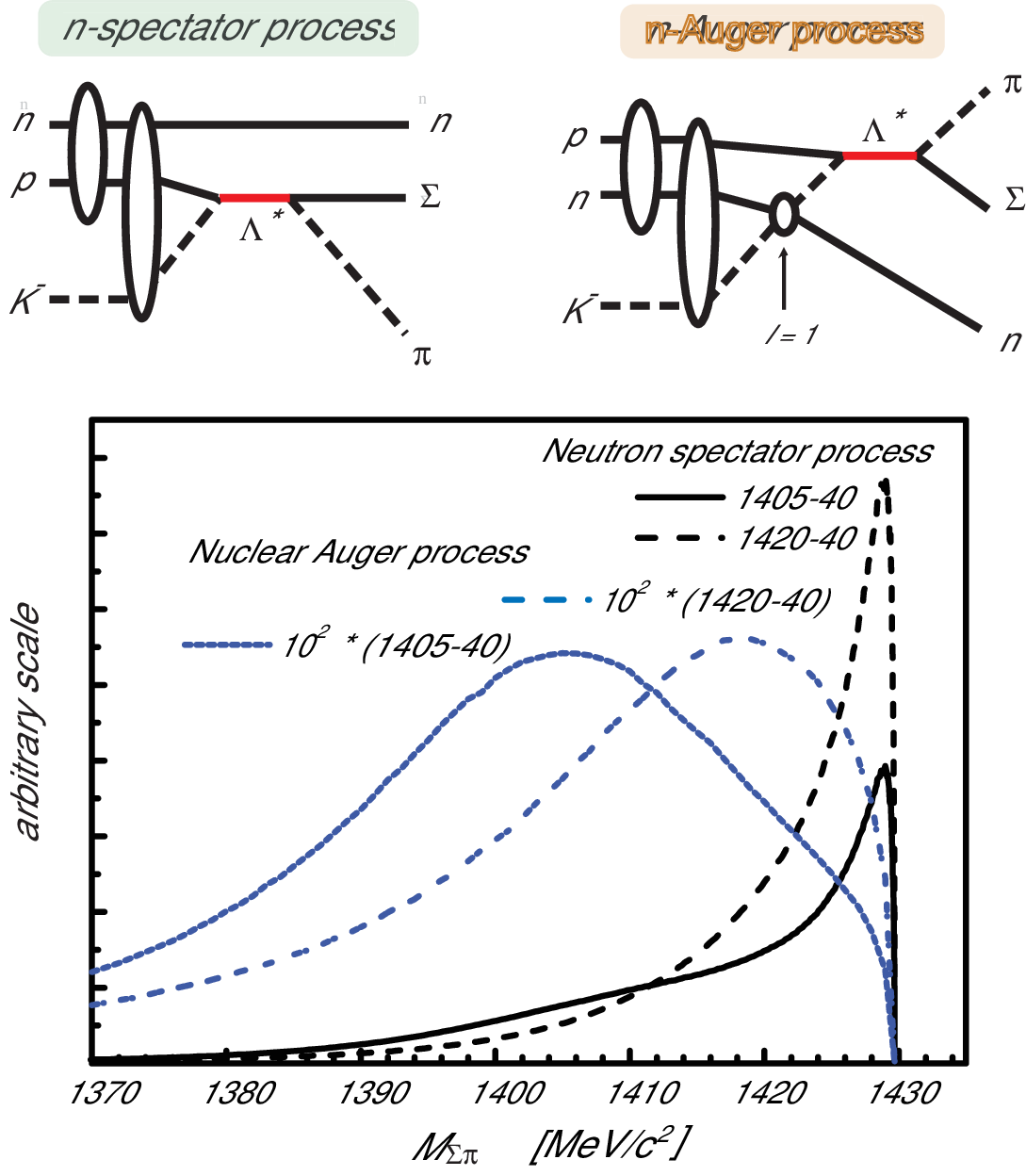}\\
  \caption{(Upper) Diagrams for the neutron-spectator and the nuclear Auger processes. (Lower)
 Calculated $M_{\Sigma\pi}$ spectra in the two processes in the case of stopped-$K^-$ on $d$.} \label{fig:Auger-process}
\end{center}
\end{figure}

\section{Coupled-channel analysis of $\Lambda(1405)$ data}

\subsection{Coupled-channel formulation of the $K^-p$ bound state}

As described in \cite{Esmaili} in detail, we treat the $K^-p$ quasi-bound state as a Feshbach resonance, considering two channels, $\bar KN$ and $\pi\Sigma$. We  employ a set of separable potentials with a Yukawa-type form factor,

\begin{eqnarray}
&& \langle \vec k'  | v_{ij}  | \vec k \rangle = g(\vec k')\,U_{ij}\, g(\vec k), \\
&& g(\vec k) = \frac {\Lambda^2} {\Lambda^2 + \vec k^2},~~U_{ij} = \frac{1}{\pi^2} \frac{\hbar^2}{2 \sqrt{\mu_i \mu_j}} \frac{1}{\Lambda} s_{ij}.
\end{eqnarray}
Here $i(j)$ stands for the $\bar KN$ channel, 1, or the $\pi \Sigma$ channel, 2, $\mu_i(\mu_j)$ is the reduced mass of channel $i(j)$, and $s_{ij}$ are non-dimensional strength parameters.

As before, we obtain $s_{11}$ and $s_{12}$ from the $M$ and $\Gamma$ values of an arbitrarily chosen $K^-p$ state to be used to calculate the $\Sigma \pi$ invariant masses. In the present coupled-channel treatment it is obvious that the properly determined two parameters, $s_{11}$ and $s_{12}$,  for any value of $s_{22}$, can represent the first pole without loss of generality. Here, we adopt $s_{22}=-0.66$, which gives $U_{22}/U_{11}= 4/3$ for $\Lambda(1405)$  as in a ``chiral" model, and $\Lambda$ = 3.90 /fm. Actually, we have proved that the obtained $M$ and $\Gamma$ do not depend on the choice of $s_{22}$ and $\Lambda$.

The coupled-channel transition matrix,
\begin{equation}
\langle \vec{k'} |t_{ij} | \vec{k} \rangle = g(\vec{k'}) \, T_{ij} \,g(\vec{k}),
\end{equation}
satisfies
\begin{equation}
T_{ij}=U_{ij}+ \sum_l U_{il} \, G_{l} \,T_{lj}
\end{equation}
with a loop function $G_l$. The solution is given in a matrix form by
\begin{equation}
T=[1-UG]^{-1}U.
\end{equation}
In our treatment, the loop function is considered to be
\begin{equation}
(UG)_{lj}=-s_{lj}\sqrt{\frac{\mu_j}{\mu_l}} \, \frac{\Lambda^2}{(\Lambda-i \,k_j)^2},
\end{equation}
where $k_j$ is the relative momentum in channel $j$.

\subsection{Observables of $\bar KN$-$\Sigma \pi$ coupled channels}

The transition matrix elements of a coupled $\bar KN(=1)-\Sigma \pi(=2)$ system are $T_{11}$, $T_{12}$, $T_{21}$ and $T_{22}$. Among them the experimentally observable quantities below the $\bar K+N$ threshold are $(-1/\pi)\,{\rm Im} \,T_{11}$, $ | T_{21} |^2\, k$ and $| T_{22} |^2\, k$, where $k$ is the $\pi \Sigma$ relative momentum. The first one, corresponding to a $\bar KN$ missing-mass spectrum, is proportional to the imaginary part of the scattering amplitude, the peak position of which is just of our concern. The second one is the $\Sigma \pi$ invariant-mass spectrum from the conversion process, $\bar KN \rightarrow \Sigma \pi$ (we call this the ``$T_{21}$ invariant mass"). The third one is a $\Sigma \pi$ invariant-mass spectrum from a scattering process, $\Sigma \pi \rightarrow \Sigma \pi$ (we call this the ``$T_{22}$ invariant mass"). 

We have derived an important relation between $T_{11}$ and $T_{21}$  from an optical relation, as given by
\begin{equation}
{\rm Im} T_{11} =  | T_{21}  |^2\,{\rm Im}\,G_2.
\end{equation}
It should be emphasized that the $T_{21}$ invariant-mass spectrum coincides with the $\bar KN$ missing-mass spectrum below the $\bar K+N$ threshold through the above formula. Thus, the observation of a $T_{21}$ invariant-mass spectrum is nothing but the observation of the imaginary part of the scattering amplitude given in Fig. 15 of Hyodo-Weise \cite{HW}.

Figure \ref{fig:MSigmaPi} (Upper) shows three observable quantities calculated with Hyodo-Weise's chiral two-channel model. The $T_{21}$ (black solid) and the $T_{22}$ (green broken) invariant masses have peaks at different positions at 1420 MeV and 1400 MeV, respectively. In comparison, the figure also shows the $T_{21}$ and the $T_{22}$ invariant masses in the $\Lambda(1405)$ ansatz of AY \cite{AY02PRC}. It is to be noted that the issue here is to discriminate whether the peak position of the $T_{21}$ spectrum is 1420 MeV or 1405 MeV. The $M_{\Sigma\pi}$ spectra in stopped-$K^- +d$ treated in the present paper are $T_{21}$ invariant-mass spectra convoluted with the deuteron momentum distribution.

\subsection{Hemingway's data and the PDG value}

The 2nd pole or resonance shape of $\Sigma\pi$ could not be obtained, unless  experimental observables involving the transition matrix $T_{22} = T_{\Sigma\pi \leftarrow \Sigma\pi}$ could be measured. The present-day PDG value of $\Lambda (1405)$ \cite{PDG} depends entirely on theoretical arguments made by Dalitz and Deloff (hereafter called DD91) \cite{DD91}. They chose exclusively 10 data points below the $\bar KN$ threshold (namely, discarding the data above the threshold) among Hemingway's $\Sigma^+\pi^-$ invariant mass spectrum \cite{Hemingway85}, and searched for the $\chi^2$ minimum for $|T_{22}|^2$ as a function of the resonance energy, $E_{\rm R}$,  under a constraint of the $I=0~\bar KN$ scattering length.  DD91 expressed a strong preference for the $M$-matrix model, and recommended a value of $(1406.5 \pm 4.0)-i \, (25 \pm 1)$ MeV, which is taken up as the PDG value.

So far, the $T_{22}$ analysis of the same experimental data of \cite{Hemingway85} has been done to materialize the double-pole hypothesis \cite{ORB,JOORM,HNJH,BNW,BMN,HW}. However, the application of $T_{22}$ to this spectrum is highly questionable, as pointed out by \cite{Oller:01}, because these $\Sigma\pi$ pairs are the decay products of $\Sigma^+ (1660)$, not those in the free scattering of $\Sigma$ and $\pi$. In fact, those who use the $T_{22}$ formula (including DD91) are destined to predict a large dip at the $\bar KN$ threshold ($M \sim 1430$ MeV/$c^2$), followed by a significant recovery above the threshold, which is characteristic of $T_{22}$. This shape disagrees seriously with the observed spectrum (see Fig.~\ref{fig:MSigmaPi}). Detailed accounts were given elsewhere \cite{Akaishi:09}.

\subsection{ANKE data}

Recently, an invariant-mass spectrum of $\Sigma^0 \pi^0$, a signature of the genuine $I=0$ $\Lambda^*$, was observed from the reaction $p + p \rightarrow K^+ + p + Y^0$ at COSY-ANKE \cite{Zychor}. We made a $\chi^2$ fitting of the spectrum by theoretical ones with the conversion $T_{21}$ involving $M$ and $\Gamma$ as parameters.
 The theoretical curves have skewed shapes, since we take into account the $\Sigma \pi$ emission threshold and the $K^- +p$ threshold realistically \cite{Akaishi:08}.
 The best-fit values we obtained are: 
$M = 1406 ^{+19}_{-9}~{\rm MeV}/c^2$ and $\Gamma = 40 \pm 8~{\rm MeV}$.
The best-fit mass does not correspond to the apparent peak position of the theoretical shape, which is found to shift downward because the broad resonance is close to the $\bar{K}N$ threshold. Although the statistical errors are large, the data are in favour of the  $\Lambda(1405)$ rather than the $\Lambda^*(1420)$ invoked by Geng-Oset \cite{Geng:07}. A future experiment with higher statistics will solve the problem.

\section{$\Sigma\pi$ invariant mass from stopped $K^-$ absorption by $d$}

 \subsection{The neutron-spectator and the nuclear Auger processes}

We showed \cite{Esmaili} that the calculated $M_{\Sigma\pi}$ spectra in $^4$He for the s-orbit capture are  favoured. Now, the situation of a liquid-deuteron target is very similar to $^4$He. When negative mesons or antiprotons are stopped in liquid-hydrogen targets, the exotic atoms formed in large-$n$ atomic orbits behave like a neutral object and undergo violent collisions, which lead to s-orbit capture after Stark mixing transitions  \cite{Stark}. Thus, we calculate $M_{\Sigma\pi}$ spectra from s-orbit absorption of $K^-$, but take into account a small contribution of the p-orbit as well. 
 
We also considered the effect of the population of the $\Sigma^0(1385)$ resonance, which is known to decay to $\Sigma^- \pi^+$ with a branching of about 5\% \cite{PDG}. It is shown that up to 20\% mixing of the $\Sigma^0(1385)$ population for $\Lambda(1405)$ and $\Lambda^*(1420)$ with $\Gamma$=40 MeV, the shapes of the spectrum are nearly unchanged. As expected, this small $\Sigma^0(1385)$ mixing can influence slightly on the  lower-mass region around 1390 MeV.

The theoretical framework for the spectator model was given in detail  in \cite{Esmaili,Yamazaki:07}. The momentum distribution of the decay particles in the $K^-$ absorption is given as:
\bea
&& \frac{d^2\Gamma}{dk_\Sigma \, dk_n} =\frac{2(2\pi)^3}{\hbar^2c^2} \mid \psi_{nlm}^{\rm{atom}}(0)
 \mid^2 \nonumber\\
&& \times \mid g(| \vec{k}_{\Sigma} + \vec{k}_n/2  |) \, T_{21}(E_2) \, g(k_n/2) \mid^2 \nonumber \\
&& \times~  k_\Sigma \, k_n\, E_\pi \mid F(k_n)\mid^2, \label{eq:Minv}
\eea

\be E_2=\sqrt{(E_i-E_n)^2-\hbar^2c^2k_n^2}-M_{\Sigma}c^2-m_{\pi}c^2,\ee
where $\psi_{nlm}^{\rm{atom}}(0)$ is a $K^-$ atomic wave function; $k_\Sigma$, $k_\pi$ and $k_n$  are the momenta of the $\Sigma$, $\pi$ and the neutron, respectively, and $|F(k_n)|^2$ is a spectator momentum distribution. The $T_{21}$ matrix involves the $\Lambda^*$ resonance effect. The kinematical constraints among the various momenta and invariant-mass spectra are given in \cite{Esmaili}.

\begin{figure}[t]
\begin{center}
 \includegraphics[width=8cm]{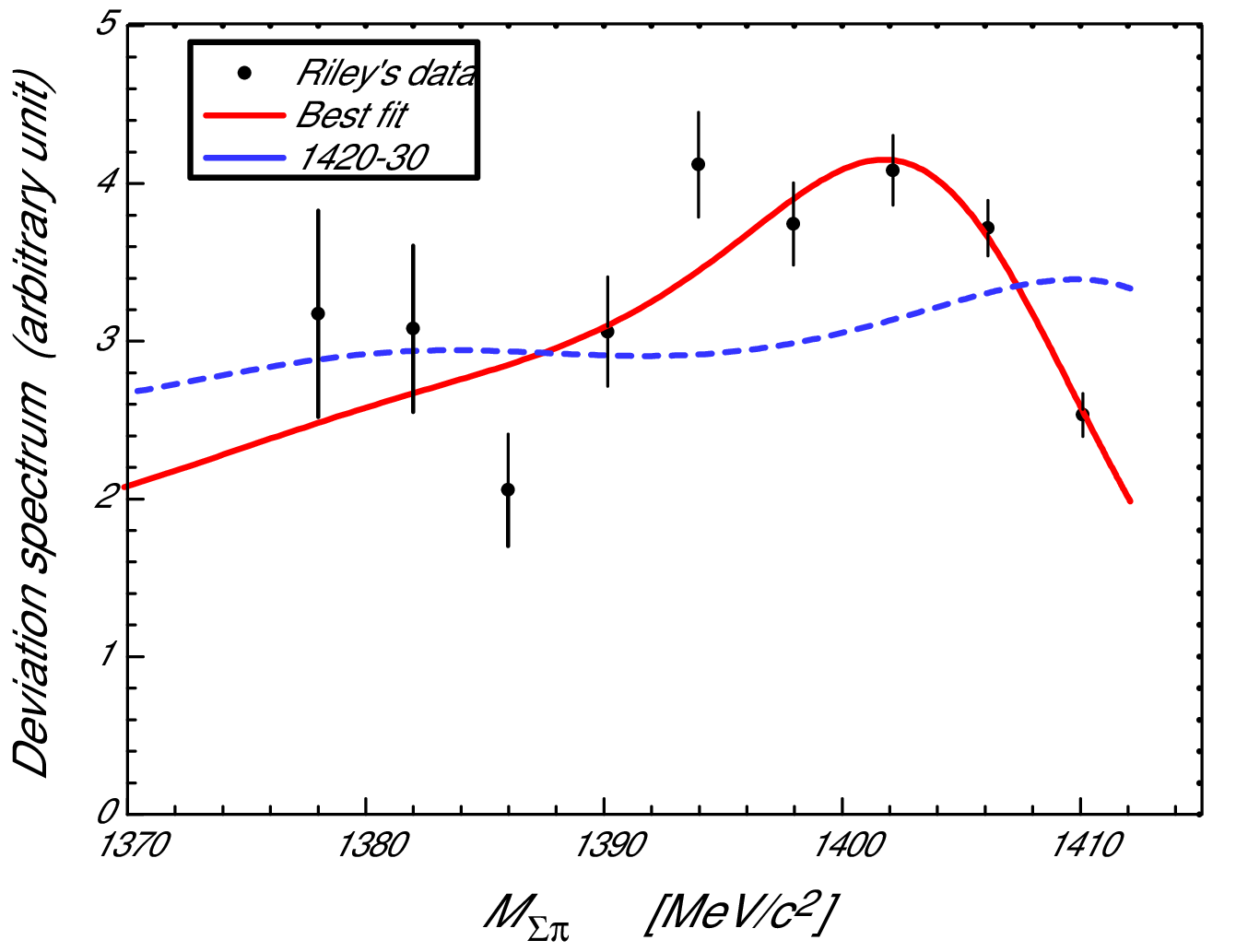}\\
 \caption{A $DEV$ expression of an experimental $M_{\Sigma\pi}$ spectrum in stopped $K^-$ in $^4$He, taken by Riley {\it et al.} \cite{Riley}. The red curve is a best-fit one with a $\chi^2 \sim 11.5$, giving $M =  1405.5^{+1.4}_{-1.0}$ MeV/$c^2$ and $\Gamma = 23.6^{+4}_{-3}$ MeV. The blue broken curve is a best-fit one with the $\Lambda^*(1420)$ ansatz, but the $\chi^2$ value is much larger ($\sim 92$).  
  }\label{fig:DEV-Riley}
 \end{center}
\end{figure}

  We have also formulated the nuclear Auger process for $K^-$ absorption at rest on $d$ similar to  \cite{Jido09}. In this process (Fig.~\ref{fig:Auger-process}), $K^-$ is scattered first by one nucleon in the $I=1$ channel, and subsequently resonant formation of $\Lambda^*$ occurs in the $I=0$ channel. Here, the neutron is not a spectator but a participant of the reaction. The neutron is kicked out from the bound orbit to the continuum with a large momentum, like an Auger electron in an atomic system. Figure~\ref{fig:Auger-process} shows a comparison of the absolute values of the $M_{\Sigma\pi}$ spectra for the neutron-spectator and the nuclear Auger processes for resonant capture to form $\Lambda(1405)$ or $\Lambda^*(1420)$.  
  
In the neutron spectator process the $M_{\Sigma\pi}$ spectrum has a QF shape with a sharp edge close to the threshold ($M = 1430$ MeV/$c^2$), which results from a small  momentum distribution of the spectator. The spectrum (\ref{eq:Minv}) is governed and projected by $|F(k_n)|^2$, because it is sharper than the resonance shape, as reflected in $T_{21} (E_2)$. Whereas the $\Lambda^*(1420)$ has an effect near the threshold, we notice that the $\Lambda(1405)$ resonance, which is located far enough from the threshold, can be populated as a nearly isolated peak.

On the other hand, in the nuclear Auger process, no such ``quasi-elastic" peak exists, and only a resonantly formed $\Lambda^*$ peak shows up. However, the nuclear Auger process has a much smaller intensity (by 2 orders of magnitude) than the neutron spectator process. It is negligible here, but becomes important in in-flight capture reactions. Thus, hereafter we adopt the neutron-spectator process as the main process to calculate the  $M_{\Sigma\pi}$ spectra.


\begin{figure}[htb]
\begin{center}
\includegraphics[width=9cm]{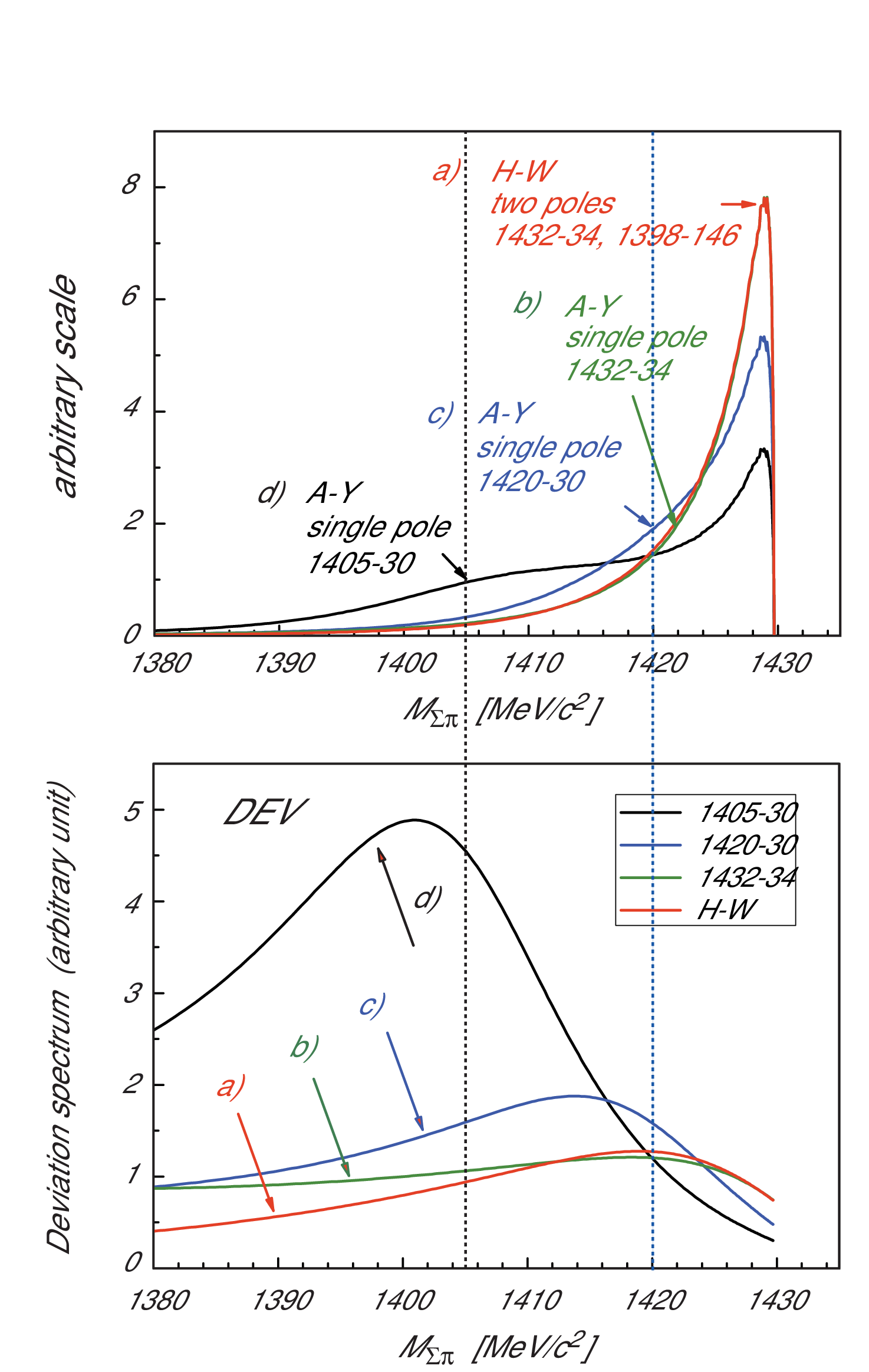}
     \caption{(Upper) $M_{\Sigma\pi}$ spectra of stopped-$K^- + d$ for different $\Lambda(1405)$ parameters including the Hyodo-Weise double-pole parameters. The spectra depend only on the 1st pole position. (Lower) $DEV$ presentation of the same $M_{\Sigma\pi}$ spectra. The assumed pole positions are indicated by vertical dotted lines.
  }\label{fig:SigmaPi-spectra}
\end{center}
\end{figure}

\begin{figure*}[htb]
\begin{center}
 \includegraphics[width=18.0cm]{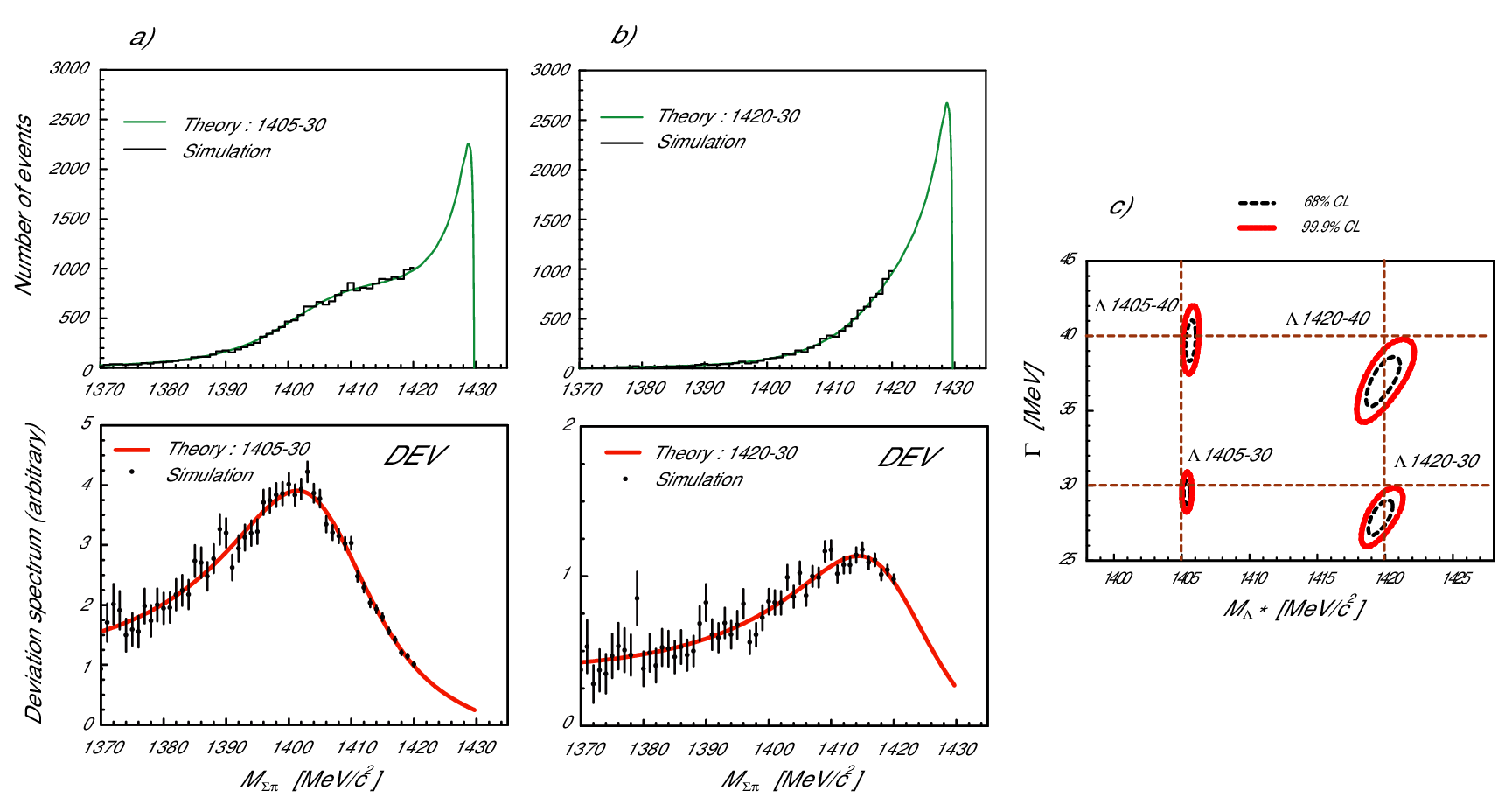}
   \caption{a) and b)  Examples of simulation spectra and corresponding $DEV$ spectra assuming $\Lambda(1405)$ and $\Lambda^*(1420)$ with $\Gamma = 30$ MeV. The spectra above 1420 MeV/$c^2$ are ``unobserved" because of the small efficiency in an actual experiment. c) Confidence contours obtained after a $\chi^2$ fitting of the four simulation spectra in the $M$-$ \Gamma$ plane.} \label{fig:simulation}
\end{center}
\end{figure*}

\subsection{$DEV$ presentation of $M_{\Sigma\pi}$ distributions}

The $M_{\Sigma\pi}$ distribution is given as follows by using Eq. (\ref{eq:Minv})
\begin{equation}
\frac{d\Gamma}{d(M_{(\Sigma\pi)^0}c^2)} = \frac{E_n}{(\hbar c)^2 k_n} \, \frac{M_{\Sigma \pi}}{M_d + m_K} 
\, \int  _0 ^{\infty} d k_{\Sigma} \frac{d^2 \Gamma}{d k_{\Sigma} d k_n}.
\label{eq:Minv2}
\end{equation}
The QF distribution without resonance is obtained by putting $T_{21} (E) \rightarrow U_{21}$.

The deuteron wavefunction is known to consist of low-momentum ($\lesssim$ 100 MeV/$c$) and high-momentum ($\gtrsim$ 100 MeV/$c$) components, as revealed in electron scattering experiments \cite{Electron}, which is well reproduced by the Av18 SC potential \cite{Argonne}. We use this realistic Av18 SC potential for the following calculations.
 The QF shape arising from the low-momentum component is narrow, whereas the high-momentum component produces a long tail in the spectrum on which resonant formation of $\Lambda(1405)$ appears. Thus, various models on $\Lambda^*$ can be easily distinguishable. Especially, in the case of our former result for $\Lambda(1405)$ (approximately represented by $M \sim$ 1405 MeV/$c^{2}$, $\Gamma \sim$ 30 MeV), the resonance state stands out as a large isolated peak, corresponding to the spectator momentum of about 170 MeV/$c$. Thus, it will be easy to prove or disprove this case. However, as the mass is closer to 1420 MeV/$c^2$, and also the width is larger, the peak separation may become unclear.

Now, we propose a new analysis method, {\it Deviation Spectrum (DEV) method}, to extract the resonance shape itself corresponding to $T_{21}$ by taking the ratio of an observed (or calculated) spectrum ($OBS$) to the non-resonant QF spectrum ($QF$) as defined by 
\be
DEV \equiv \frac{OBS ~({\rm observed~or~calculated})} {QF~({\rm non}~{\rm resonant})}.
\ee
It is to be noted that the ``$QF$"  in the denominator, given by Eq.  (\ref{eq:Minv}, \ref{eq:Minv2}) with $T_{21} (E) \rightarrow U_{21}$, is a scaling function common to any observed spectrum as well as to any model spectrum. Thus, we can see that
\begin{equation}
DEV \propto | T_{21} (E_2)  |^2, ~ E_2 = (M_{\Sigma \pi} -M_{\Sigma} - m_{\pi} ) c^2.
\end{equation}
A $DEV$ spectrum is essentially equivalent to its original $M_{\Sigma\pi}$ distribution, and has a practical merit that the resonant component, $|T_{21} (E_2)|^2$, is extracted as a  visualized form so that one can notice its presence even {\it by eyes without computer fitting}. One can also recognize that the existence of a resonance peak itself is not affected much by small ambiguities in the $QF$ shape. 

As an example we express in Fig.~\ref{fig:DEV-Riley} old Riley's data of $M_{\Sigma\pi}$ from the stopped $K^-$ reaction in $^4$He \cite{Riley} in a $DEV$ spectrum. Although this data was once analyzed in our previous paper, yielding a meaningful result, (\ref{eq:Riley-results}), the presence of the $\Lambda(1405)$ resonance was hardly visible. Now, this DEV plot shows clearly a peak-like structure. The best-fit curve (red) with a $\chi^2 \sim 11.5$ for $ndf =  6$ gives the same results as the previous values without $DEV$ before, (\ref{eq:Riley-results}).  On the other hand, the $\Lambda^*(1420)$ ansatz gives a large deviation from the $DEV$ points ($\chi^2 \sim 92$ for $ndf =  8$). The stopped-$K^-$ on $^4$He is a rather difficult case because the QF peak is broader, but the $DEV$ method proves to work well. The case of a deuteron target is more straightforward, as shown below.  
 
We calculated the $M_{\Sigma\pi}$ spectra and their corresponding $DEV$ spectra for stopped $K^-$ on $d$ with s-orbit absorption for various models and parameters of $\Lambda(1405)$. Figure~\ref{fig:SigmaPi-spectra} shows the single-pole cases with parameters of $M$-$\Gamma$ = 1405-30 and 1420-30. In addition, the calculated spectrum shape based on the double-pole model of Hyodo-Weise (1432-34 and 1398-146) is shown together with the single-pole case of 1432-34. The double-pole case is found to be in almost exact agreement with the corresponding single-pole case, clearly indicating that the 2nd pole has no visible effect on $M_{\Sigma\pi}$. We also examined this point using a model of R$\acute{\rm e}$vai-Shevchenko \cite{Revai}, which provides both single-pole and double-pole cases. There is virtually no difference between the two spectra, indicating that the 2nd pole is irrelevant to the experimental observables.

\begin{figure}[htb]
\begin{center}
\includegraphics[width=6cm]{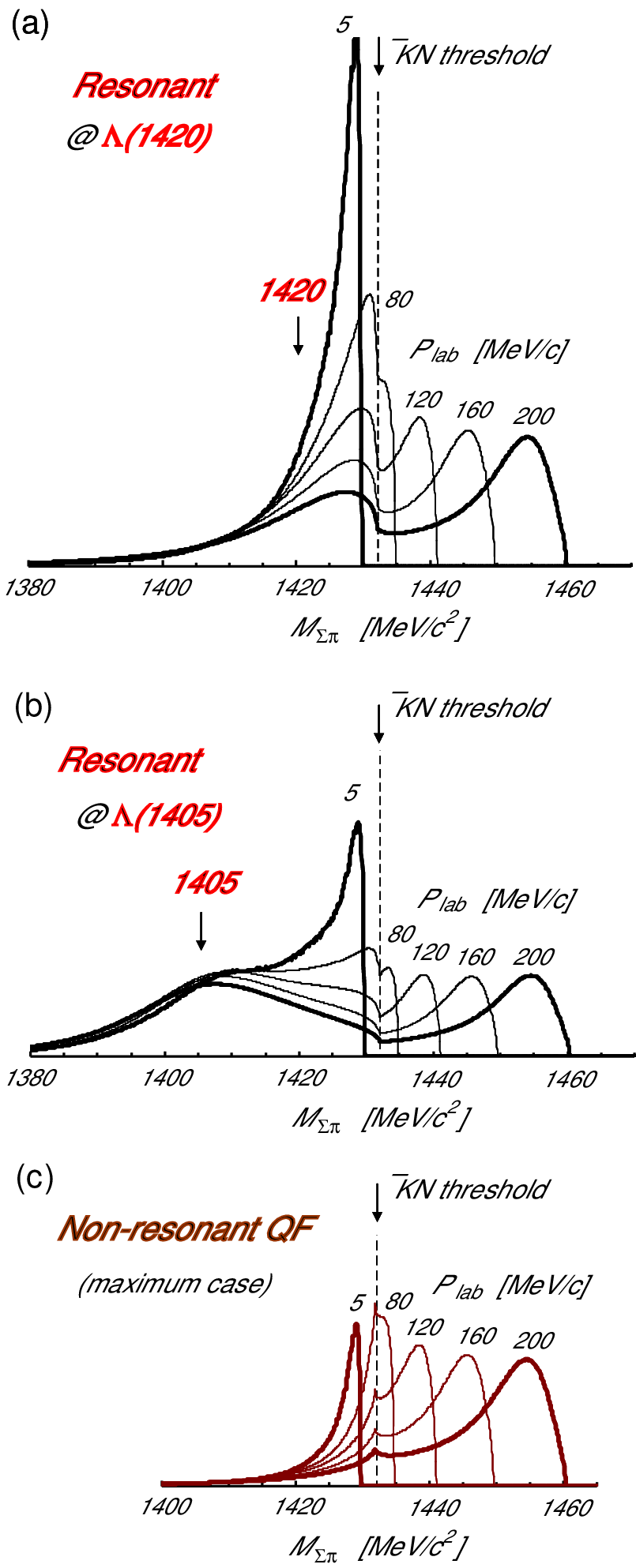}\\
 \caption{$T_{21} ~\Sigma \pi$ invariant-mass spectra from low-momentum $K^- + d$ reactions calculated with AY interactions. The total spectra (resonant plus non-resonant): (a) for AY 1432-34 and (b) for AY 1405-30. The curves (a) and (b) are normalized to unit area to compare with $K^-d$ atomic absorption. (c) The case of non-resonant QF contribution for (a), in which the resonant QF contributions via $\Lambda^*$ are subtracted by reducing $U_{11}$ to $0.59 \, U_{11}$ where the $\Lambda^*$ quasi-bound state disappears.}
 \label{fig:QF}
 \end{center}
\end{figure}

\subsection{Data simulation and fitting}

An experiment is planned at J-PARC \cite{Suzuki:09,Suzuki-JPARC}. The invariant-mass spectrum can be obtained as a missing-mass spectrum of $\Delta M (Kd,n)$ from observations of $n$ together with $\Sigma\pi$ events. The mass resolution depends on the momentum resolution of $n$. It is expected to be better than 2 MeV ($\sigma$), good enough for the present purpose (reconstruction of $M_{\Sigma\pi}$ from $\Sigma$ and $\pi$ momenta gives a much worse resolution). However, the neutron detection efficiency drops down for the mass region above 1420 MeV/$c^2$. This means that the quasi-free peak region is missing from observations. To examine whether we can obtain useful information on the mass and width of $\Lambda^*$ from an experimental missing-mass spectrum, we produce simulation spectra assuming the cases of $\Lambda(1405)$ and $\Lambda^*(1420)$, both with $\Gamma = 30$ MeV, as shown in Fig.~\ref{fig:simulation} a, b), respectively.

The $DEV$ spectra of simulation events are shown in Fig.\ref{fig:simulation}. For each case the $DEV$ points alone reveals a resonance shape even without fitting. We attempt to fit the simulation spectra by a theoretical spectrum with ($M, \Gamma$) as arbitrary parameters, and draw $\chi^2$ confidence contours in Fig.~\ref{fig:simulation} c). Although the simulation spectra lack the region above 1420 MeV, the fitting results are remarkable. The data a) ($\Lambda$1405-30) produce fitting results: $M = 1405.4 \pm 0.3$ and $\Gamma = 29.6 \pm 0.8$, which are in excellent agreement with the input data. On the other hand, the data b) ($\Lambda$1420-30) give $M = 1419.6 \pm 1.0$ MeV/$c^2$ and $\Gamma = 27.9 \pm 1.2$ MeV. The fact that the $\Gamma$ value deviates slightly from the input data can be understood in view of the missing region ($M > 1420$ MeV/$c^2$) of the simulation spectrum. Nevertheless, the present fitting procedure turns out to be capable of extracting the value of $M$ to high precision. It is to be emphasized that the $\Lambda(1405)$ ansatz can be proven or disproven from an observed $DEV$ spectrum in a straightforward way.

\section{In-flight $K^- + d$ reactions for $\Lambda(1405)$}

 Recently Jido, Oset and Sekihara (JOS) \cite{Jido09} calculated in-flight $K^- + d$ reactions for $\Lambda(1405)$, in which they stated as follows:  in the stopped $K^-$ case ``the impulse approximation
term is absolutely dominant and the trace of the $\Lambda(1405)$ is lost", and ``one can conclude
that the case of stopped kaons does not provide a good set up to learn about the $\Lambda(1405)$".
Now, we examine this statement. Figure \ref{fig:QF} shows quasi-free (QF) contributions calculated with AY interactions for $K^- + d \rightarrow \Sigma + \pi + n$ reactions of $n$-spectator process with incident $K^-$ momenta of 5, 80, 120, 160 and 200 MeV/$c$. The figures (a) and (b) are the total spectra for the $\Lambda^*(1420)$ and $\Lambda(1405)$ cases, each of which includes both the {\it resonant} ($K^- + "p"~{\rm in}~ d \rightarrow \Lambda^* \rightarrow \Sigma + \pi$) and {\it non-resonant direct} contributions. We try to separate the two contributions.

The $T_{21}$ matrix element can be written in the form with final-state interaction (FSI) and initial-state interaction (ISI) \cite{Akaishi:09},
\begin{eqnarray}
T_{21} &=& \frac{1}{1-U_{22}^{\rm {opt}} G_2} U_{21} \frac{1}{1-U_{11} G_1} \nonumber \\
&=& f_{\rm {FSI}}(U_{22}^{\rm {opt}})~ U_{21}~ f_{\rm {ISI}} (U_{11}),
\label{ISI}
\end{eqnarray}
where
$$ U_{22}^{\rm {opt}} = U_{22} + U_{21} \frac{G_1}{1-U_{11} G_1} U_{12}. $$
If we change the strength of the $\bar KN$ interaction as $U_{11} \rightarrow f \cdot U_{11}$, the $\Lambda^*$ quasi-bound state (QBS) disappears at $f=0.59$ in the case of (a). At $P_K =  5$ MeV/$c$ the QF strength is reduced to 21\%, 7.9\% and 3.9\% of the Fig.~\ref{fig:QF}(a) value for $f = 0.59$, 0.3 and 0.0, respectively: this change is mainly due to ISI. The non-resonant QF is estimated to be in between 21\% (no QBS) and 3.9\% (no ISI) which is enhanced by FSI from the Born term (the $U_{21}$ term alone) of 0.67\%.  The case of Fig.~\ref{fig:QF}(c) shows the maximum contributions of the non-resonant QF processes for the case (a), where a {\it cusp structure appears at the $K^-+p$ threshold}. The observable spectrum is of the case (a) or (b), where the {\it resonant QF contribution undoubtedly dominates} with 80-95\% of the total QF strength in the quasi-bound region. Although JOS took the same resonant QF process into account and obtained essentially the same result as ours, they simply missed the dominance of the resonant formation in the QF process and misinterpreted their spectrum as if it were in the non-resonant QF case (c), leading to a  wrong conclusive statement on the effectiveness of the stopped $K^-$ method.

Needless to say, the $\Lambda^*$ issue can be examined by an in-flight $K^-$ experiment as well, but the effectiveness of the stopped-$K^-$ method is generally superior from various experimental points of view, such as the reaction rate and the achievable mass resolution.

\section{Conclusion \label{conclusion}}

We have shown that the issue of $\Lambda(1405)$ vs $\Lambda^*(1420)$ can be solved from experimental observables of the conversion $M_{\Sigma\pi}$, for which $T_{21}$ is responsible. Conrary to the prevailing belief the effect of the presence of a 2nd pole, if any, is negligible in such spectra. Since a deuteron has both low-momentum and high-momentum components rather separately, the $M_{\Sigma\pi}$ spectral shape from stopped-$K^-$ absorption in $d$ is characterized by a narrow direct-capture component, followed by a long tail, on which a resonantly formed $\Lambda^*$ will be revealed somewhere in a wide mass region. We have proposed an efficient method of taking deviation spectra ($DEV$) to extract the resonant shape of $|T_{21}|^2$. The case of $\Lambda(1405)$ will be easily proven or disproven. A proposed  experiment at J-PARC will yield a decisive result on the current debate.

\section{Acknowledgements}

The authors are grateful to Dr. T. Suzuki for his stimulating discussion. One of us (J.E.) would like to thank Professors E. Hiyama and M. Iwasaki for the hospitality at RIKEN, and also Professor S.Z. Kalantari for helping him before coming to Japan. He is also grateful to Nishina Memorial Foundation for the receipt of Nishina Memorial Fellowship.  This work is supported by the Grant-in-Aid for Scientific Research of Monbu-Kagakusho of Japan.

\newpage

\end{document}